\documentclass[10pt, conference]{IEEEtran}
\IEEEoverridecommandlockouts
\usepackage{cite}
\usepackage{amsmath,amssymb,amsfonts}
\usepackage{algorithmic}
\usepackage{textcomp, multirow}
\usepackage{color}
\usepackage{subfig}

\def\BibTeX{{\rm B\kern-.05em{\sc i\kern-.025em b}\kern-.08em
    T\kern-.1667em\lower.7ex\hbox{E}\kern-.125emX}}

\ifCLASSINFOpdf
  \usepackage[pdftex]{graphicx}
  \graphicspath{{./}}
  \DeclareGraphicsExtensions{.pdf,.jpg,.png,.xps}
\else
\fi

\IEEEoverridecommandlockouts
\IEEEpubid{\makebox[\columnwidth]{978-1-5090-0223-8/16/\$31.00~\copyright~2016 European Union \hfill} \hspace{\columnsep}\makebox[\columnwidth]{ }}
\begin{document}

\title{Improving Reliability of Service Function Chains with Combined VNF Migrations and Replications}

\author{\IEEEauthorblockN{Francisco Carpio and Admela Jukan}
	\IEEEauthorblockA{Technische Universit{\"a}t Braunschweig, Germany}
	\IEEEauthorblockA{Email:\{f.carpio, a.jukan\}@tu-bs.de}
}

\maketitle

\begin{abstract}
The Network Function Virtualization (NFV) paradigm is enabling flexibility, programmability and implementation of traditional network functions into generic hardware, in form of Virtual Network Functions (VNFs). To provide services, the VNFs are commonly concatenated in a certain ordered sequence, known as Service Function Chains (SFCs). SFCs are usually required to meeting a certain level of reliability. This creates the need to place the VNFs while optimizing reliability jointly with other objectives, such as network and server load balancing. Traditional migration and replication mechanisms, commonly used for Virtual Machines (VM) in data centers, can be used to improve SFC reliability. We study how to improve service reliability using jointly \emph{replications} and \emph{migrations}, considering the chaining problem inherent in NFV. While replications provide reliability, performing migrations to more reliable servers decreases the resource overhead. A Linear Programming (LP) model is presented to study the impact of active-active configurations on the network and server resources. Additionally, to provide a fast recovery from server failures, we consider N-to-N configurations in NFV networks and study its impact on server resources. The results show that replications do not only improve reliability, but can also be used to achieving a better server and network load balancing, and when used jointly with migrations can improve resource utilization without degrading reliability.
\end{abstract}


\section{Introduction}

Network Function Virtualization (NFV) is a new paradigm that virtualizes the traditional network functions and places them into generic hardware inside the network or in data centers, as opposed to the traditional designated hardware. Since a single Virtual Network Functions (VNF) cannot provide a full service, multiple VNFs are commonly linked in a sequence order, known as Service Function Chains (SFCs), and placed into the network which introduces the so-called VNF placement problem. The placement of the VNFs can happen either in data centers (DC) or by deploying standard Physical Machines (PMs) (e.g., commodity servers) inside the network. Those SFCs provide services that using quality metrics require to meet certain key performance indicators for reliability, latency, service outage downtime, etc.  

In NFV-based networks, the provision of service reliability is critical because the failure of a single VNF breaks the continuity of the SFC. Replication mechanisms have already been proposed to target the required service reliability based on VNF redundancy,  which allow configurations in Active-StandBy or Active-Active modes. The Active-StandBy configurations instantiate VNF replicas in the network without providing service, just waiting for some failure on the main VNF to start working. These configurations do not require any load distribution function, since the replicas are not performing any task during the normal operation mode. On the other hand, Active-Active configurations instantiate VNF replicas which are fully functional VNFs, even in the normal operation mode. In this mode, when a VNF fail the affected traffic is redirected to the remaining operational VNFs. To achieve that, a load distribution function is required as well as a coordination function to maintain the internal state among the replicas. 

This paper studies the impact of active-active configurations on the network and server utilization. The main objective is to improve reliability of the service chains by a combined allocation of replicas and usage of migrations of VNFs from lower to higher reliable servers. Since active-active configurations in NFV standards do not consider the reservation of resources to minimize the reconfiguration time in case of failures, some solutions propose traffic replication to avoid reconfiguration delays. In contrast, we propose to study the impact on the server resources when N-to-N configurations are used in NFV. Those schemes consider the reservation of resources before failures to minimize the reconfiguration times, but without the need of replicating traffic. To this end, we propose an LP model to optimally perform placement, replications and migrations of VNFs while maximizing the service reliability and minimizing required network and server resources. 

The rest of the paper is organized as follows. Section II presents related work. Section III describes the proposed solution. Section IV formulates the optimization model. Section V analyzes the performance and Section VI concludes the paper.

\section{Related Work}

Different methods to provide end-to-end reliability in NFV environments has already been discussed by ETSI \cite{ETSI2016}. The proposed schemes for VNF protection can be classified in two main groups: Active-Standby and Active-Active methods. Active-Standby is a straightforward solution  where it is not required to have load distributions functions, but a mechanism to redirect traffic towards standby nodes in case of failure. This solution requires to reserve dedicated resources that are not performing any task. In contrary, active-active schemes are not standby entities, but all nodes are providing service. This solution not only requires redirection of traffic in case of failure, but also an additional load balancing function before the pool of replicated entities to decide how to distribute the traffic. The complexity of active-active solutions is in maintenance and coordination of the internal state replication among replicas to ensure the stateful sessions. 

Previously proposed redundancy methods have considered the physical hardware reliability, while ignoring the global information of the VNF Forwarding Graph (VNF-FG) and leading to inaccurate estimation for the services. This has led to an inefficient utilization of networking resources, as shown in \cite{Ding2017}. Therefore, the service reliability became rather important in NFV networks. For instance, in \cite{Kang2017}, the authors evaluate the reliability criterion within a probabilistic model and propose an LP model to address the joint optimization of the chain composition (CC) and the FGE problem. The same problem is solved with heuristics by \cite{Ye2016} and \cite{Liu2016}. This body of work concluded that an optimum mapping of VNFs is required to optimize reliability regardless the necessity to instantiate additional VNF replicas. Similarly, \cite{Hmaity2016} showed that to guarantee resiliency against single-node failures, it was required the duplicate the amount of resources even in case of failure of a single VNF in service chains.

In the meantime, data centers are being the first environment where to deploy VNFs due to the high flexibility provided by cloud operators. In \cite{Herker2015}, the authors model different SFC backup strategies and provide heuristic algorithms for resilient embedding in different data center topologies analyzing which one provides a better throughput for a certain resilience. To improve reliability, one solution is to place VNFs backups on the routing path that the SFC is using, however this increases the end-to-end delays due to the extension of the path. To solve that problem, \cite{Qu2017} proposed multipath backup schemes to increase reliability while, at the same time, minimizing end-to-end delays. The proposed LP model also replicates the traffic between two end-points in order to support immediate recovery after a failure. While applying this solution does not imply disruption on the service, redundant network and server resources are necessarily to make this solution work. 

We showed in our previous work \cite{Carpio2016a} that VNF placement with replications can load balance network, and in this paper we add reliability and server usage optimization in the same type of system. Replicas generally utilize additional resources, we combine replicas with migrations, from lower reliability to higher reliability servers. Additionally, since resource reservation method is required in NFV networks to provide fast recovering after server failures, we also provide a study of the impact of N-to-N configurations have on the server utilization.

\begin{figure}[!t]
\centering
\subfloat[Scheme without protection]{
	\label{subfig:single}

	\includegraphics[width=0.39\textwidth]{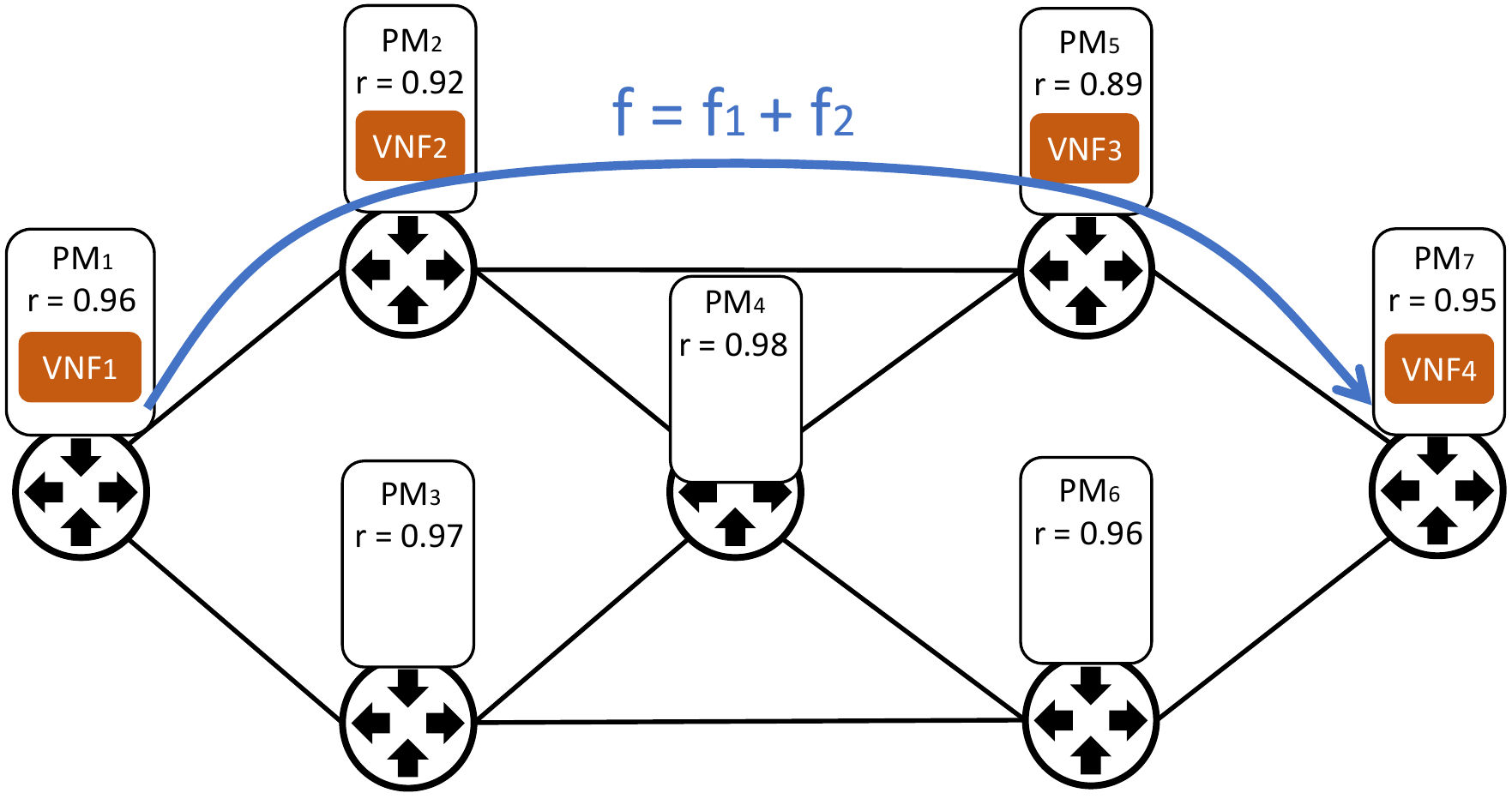} } 

\subfloat[Improving reliability with migration of VNFs]{
	\label{subfig:migration}
	\includegraphics[width=0.39\textwidth]{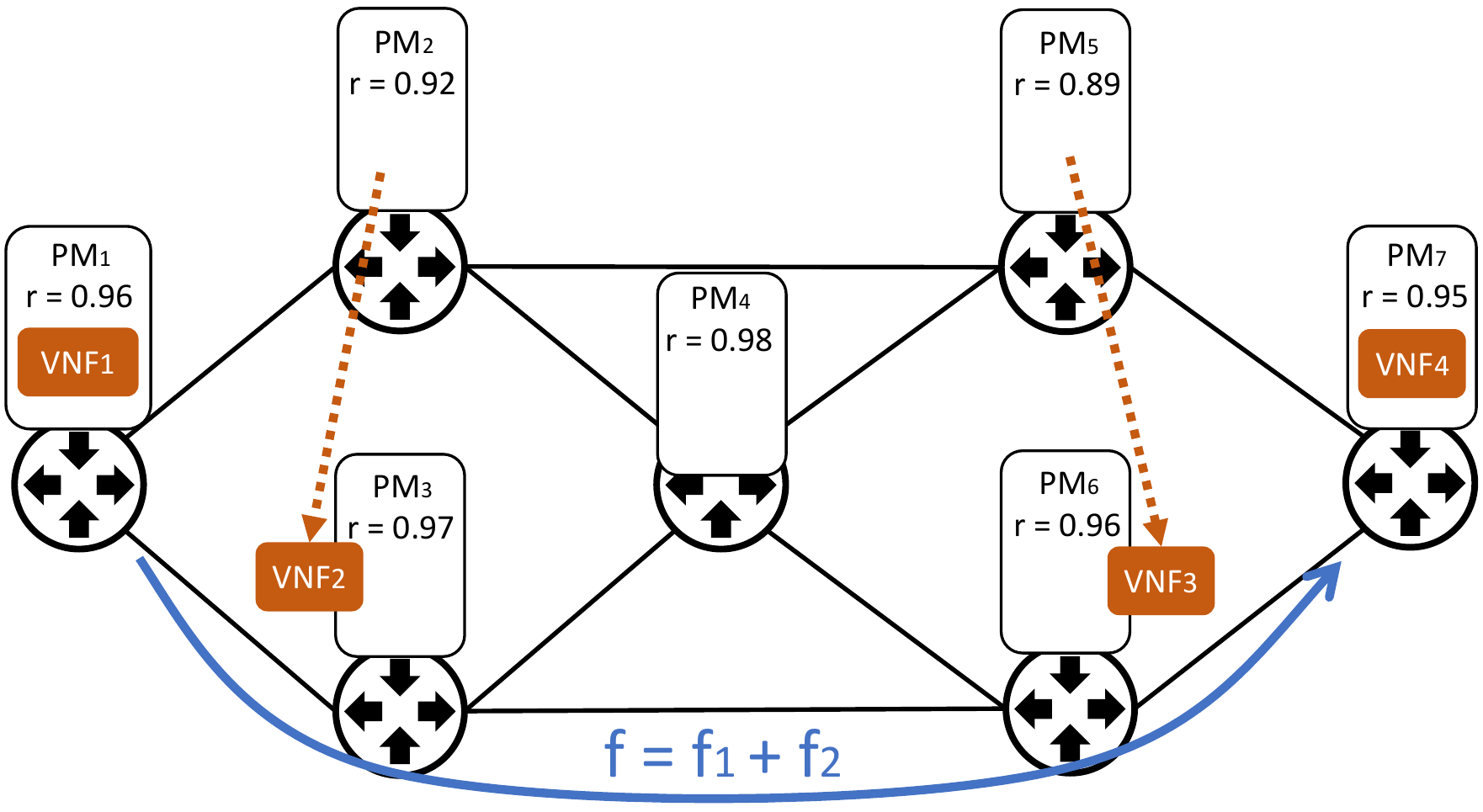} } 

\subfloat[Active-Active protection scheme]{
	\label{subfig:replication}
	\includegraphics[width=0.39\textwidth]{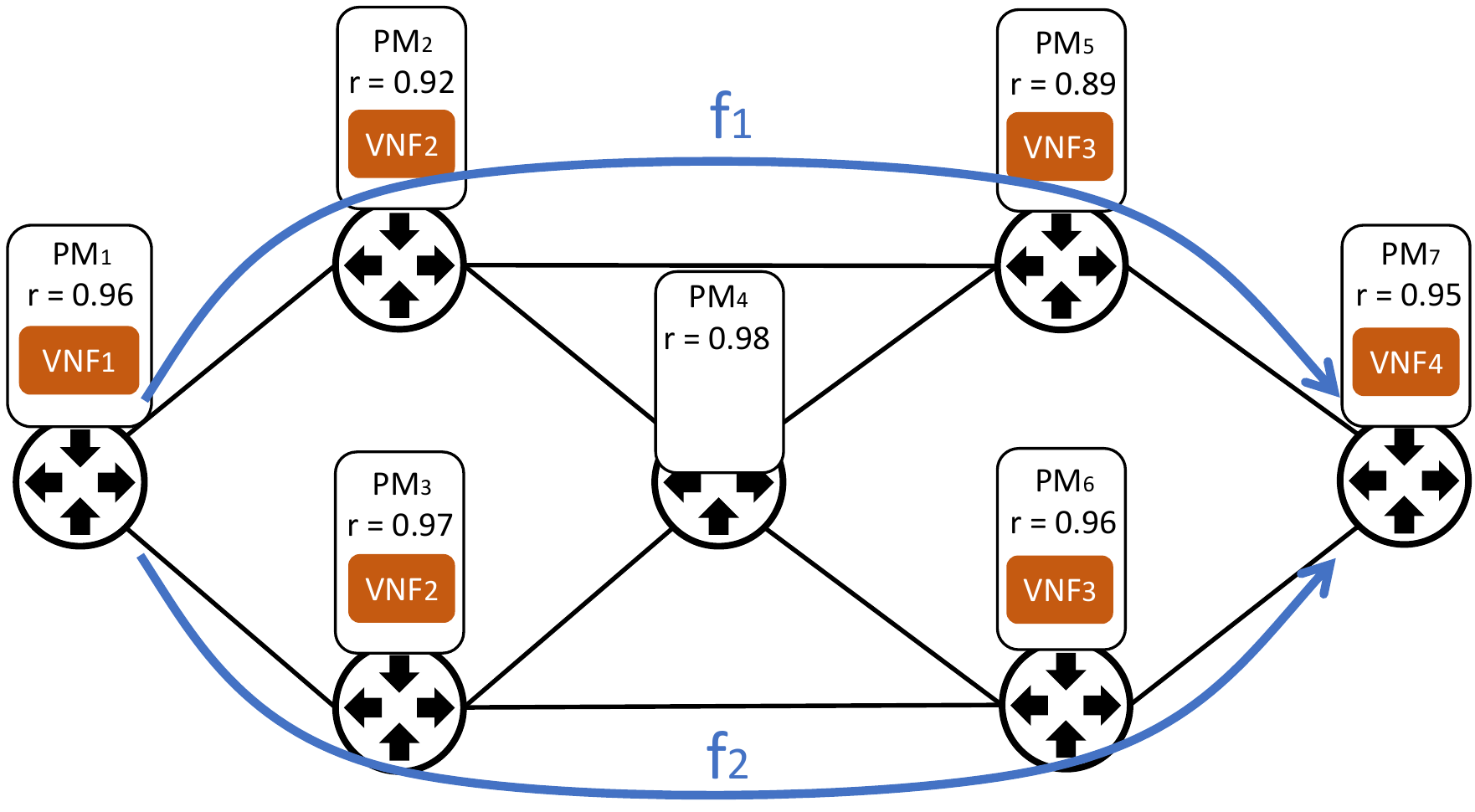} } 

\subfloat[Active-Active protection scheme with migration of VNFs]{
	\label{subfig:hybrid}
	\includegraphics[width=0.39\textwidth]{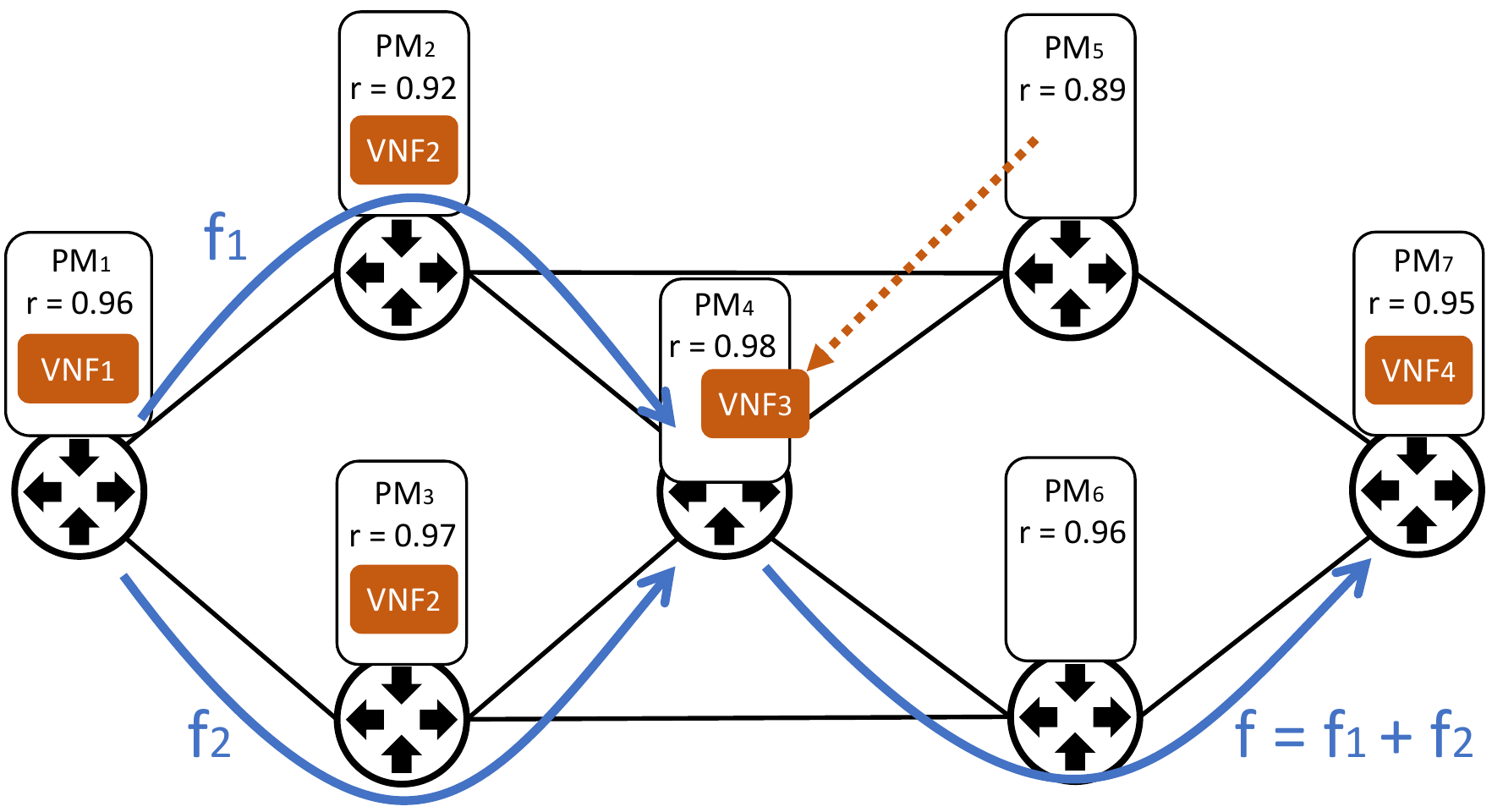}} 
\caption{Methods to improve reliability of SFCs}
\vspace{-0.4cm}
\label{methods}
\end{figure}

\section{Reliability Solutions in NFV networks} 

In this section, we discuss the existing methods to provide reliability with migrations and replications, and introduce our proposed solution that combines the two to provide even better reliability of service function chains. Let $S$ be the set of service chains and $V_s$ the set of VNFs from a specific service chain $s$. Then, according to \cite{ETSI2016}, the reliability of a certain service chain is calculated as:
\begin{equation} 
	\forall s \in S: R_s = \prod_{v \in V_s} R_v
\end{equation}
, where $R_v$ is the reliability of a VNF $v$. Following the example shown in Fig. \ref{subfig:single}, where 4 concatenated VNFs are allocated in $PM_1$, $PM_2$, $PM_5$ and $PM_7$, respectively, the reliability of the service chain is: $R_s = 0.96 \cdot 0.92 \cdot 0.89 \cdot 0.95 = 0.747$. In case the reliability does not meet the requirements, one solution is to migrate the VNFs to another $PMs$ with higher reliabilities, as shown in Fig. \ref{subfig:migration}, where $VNF_2$ and $VNF_3$ are migrate to $PM_3$ and $PM_6$, respectively. In this case, the achieved reliability is: $R_s = 0.96 \cdot 0.97 \cdot 0.96 \cdot 0.95 = 0.849$. If the reliability is still lower than the minimum requirements, the solution goes through instantiating VNF replicas in order to improve the robustness. Then, the reliability is defined as:
\begin{equation} \label{reliability_equation}
	\forall s \in S: R_s = \prod_{v \in V_s} \bigg[ 1 - \prod_{w \in V_{v}^s} (1 -R_w) \bigg]
\end{equation}
, where $V_{v}^s$ is the original VNF $v$ plus the set of replicas of the same function. An example is shown in Fig. \ref{subfig:replication}, where the $VNF_2$ and $VNF_3$ are replicated into $PM_3$ and $PM_6$, respectively. In this case, the achieved reliability is: $R_s = 0.96 \cdot [1 - (1 -0.92)(1 - 0.97)] \cdot [1- (1-0.89)(1- 0.96)] \cdot 0.95 = 0.906$. The more replicas are instantiated, the higher becomes the reliability of the service chain. This solution, known as active-active \cite{ETSI2016}, allocates multiple instances of the same VNF running in active-mode. This scheme requires a load balancing function in front of the pool of replicas to load balance the traffic. In case of failure of one of the replicas, the service chain is not interrupted at all. Only the traffic is redirected from the failed instance to the remaining operational ones. 

The main inconvenience with active-active schemes is that higher the reliability achieved, larger the amount of the required resources. For that reason, it is important to consider the optimum allocation of VNFs, from a global point of view, as a requirement for resilient NFV networks. In order to minimize the number of required replicas, while maintaining a similar level of reliability, we propose to use \emph{migrations jointly with replications} of VNFs. An example is shown in Fig. \ref{subfig:hybrid}. Here, only the $VNF_2$ is replicated into the $PM_3$ while the traffic is also splitted. Then, in case of failure of $PM_2$, the $f_1$ would be redirected towards $PM_3$ without service interruption, only the delay associated to the  redirection of the traffic. Differently from the previous example, here the $VNF_3$ is migrated into the $PM_4$, instead of being replicated, allowing the increment of reliability without using extra resources. The achieved service reliability in this case is $R_s = 0.96 \cdot [1 - (1 -0.92)(1 - 0.97)] \cdot 0.98 \cdot 0.95 = 0.892$, which is slightly lower than in the previous case, but requiring one VNF less. 

\begin{figure}[!t]
	\centering
	\includegraphics[width=0.8\columnwidth]{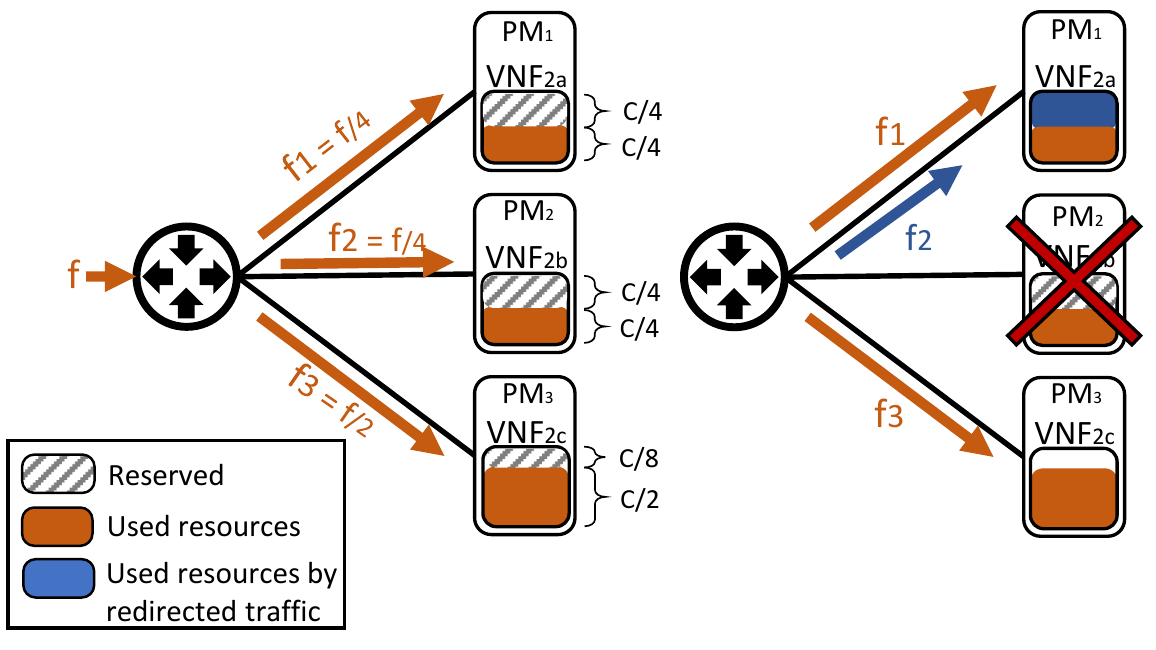}
	\caption{N-to-N protection scheme}
	\label{operation}
\vspace{-0.5cm}
\end{figure}

However, the lack of reservation of resources to accommodate the extra redirected traffic from the failed nodes, makes the remaining operational nodes to require more time to accommodate the redirected traffic. To solve that problem, N-to-N configurations, which are a combination between active-active and N+M configurations, reserve extra resources for fast recovering in each one of the active nodes, without the necessity to have dedicated nodes in a standby mode. This scheme is shown in Fig. \ref{operation}. Here, the $VNF_{2a}$, $VNF_{2b}$ and $VNF_{2c}$, which are replicas of the same function, are processing three different traffic flows, $f_1, f_2$ and $f_3$ , respectively. Because $f_3$ is twice the size of $f_1$ and $f_2$, the used resources in $PM_3$ are also twice than those used in $PM_1$ and $PM_2$. In order to support the redirected traffic in case of one of the PMs fails, each $VNF$ requires to reserve extra capacity. Assuming $C$ as the capacity of a PM and $f$ the maximum traffic processed by one PM, for instance, the reserved capacity in $VNF_{2a}$ and $VNF_{2b}$ could be $C/4$ in order to support half of the redirected flow $f_3$, in case of failure. While in the $VNF_{2c}$ could be only $C/8$ to support half of the redirected flows $f_1$ or $f_2$. With this configuration there are many possibilities, for instance, in case of failure of $PM_2$, the flow $f_2$ can be completely redirected towards the $PM_1$ or can be split and redirected to both $PM_1$ and $PM_3$. Which one of these solution is chosen will depend whether the objective is to load balance the servers, the network, or both. 
 
In this paper, we propose to deploy both replications and migrations of VNFs to increase the reliability of services, but minimizing the impact on the required network and server resources. Additionally, we also propose to study the impact that N-to-N configurations have on the server utilization.

\begin{table}[!t]
	\renewcommand{\arraystretch}{1.3}
	\caption{Notation}
	\label{parameters}
	\centering
	\footnotesize
	\begin{tabular}{c p{6.5cm}}
		\hline
		\textbf{Parameters} & \textbf{Meaning}\\
		\hline
		$N, X, L$ & set of nodes, servers (i.e. PMs) and links, respectively\\
		$S, V_s, \Lambda_s$ & set of all service chains, ordered set of VNFs in service chain $s$ and set of traffic demands per service chain $s$, respectively\\
		$R, Y, P$ & set of reliabilities, linear cost functions and all pre-computed paths, respectively\\
		$P_s \subseteq P$ & set of available paths for service chain $s$\\
		$N_p \subseteq N$ & ordered set of nodes traversed by path $p$\\
		$X_n \subseteq X$ & set of servers in node $n$\\
		$R_x \subseteq R$ & reliability of server $x$\\
		$T_{p}^\ell$ & 1 if path $p$ traverses link $\ell$\\
		$F_{v}$ & 1 if function $v$ can be replicated\\
		$F_{MAX}$ & maximum number of allowed replicas per service chain\\
		$L_v$ & load ratio of VNF $v$\\
		$E_m , E_r$ & penalty ratio due to migration, replication\\    
		$C_\ell, C_x$ & maximum capacity of link, server \\
		\hline
                \textbf{Variables} & \textbf{Meaning}\\
                \hline
                $t_{p}^s$ & A binary routing variable, 1 if service chain $s$ is using path $p$ \\
		$t_{p}^{\lambda,s}$ & A binary routing variable, 1 if traffic demand $\lambda$ from service chain $s$ is using the path $p$\\
		$f_x^{v,s}$ & A binary variable, 1 if VNF $v$ from service chain $s$ is allocated at server $x$\\
		$f_{x,\lambda}^{v,s}$ & 1 if VNF $v$ from service chain $s$ is being used at server $x$ by traffic demand $\lambda$\\
		$k_\ell, k_x, k_v, k_s$ & utilization cost of link $\ell$, utilization cost of server $x$, migration cost of VNF $v$ and reliability cost of a service chain $s$, respectively\\
		\hline
		\vspace{-0.6cm}
	\end{tabular}
\end{table}

\section{Problem Formulation}

\begin{figure}[!t]
	\centering
	\includegraphics[width=0.7\columnwidth]{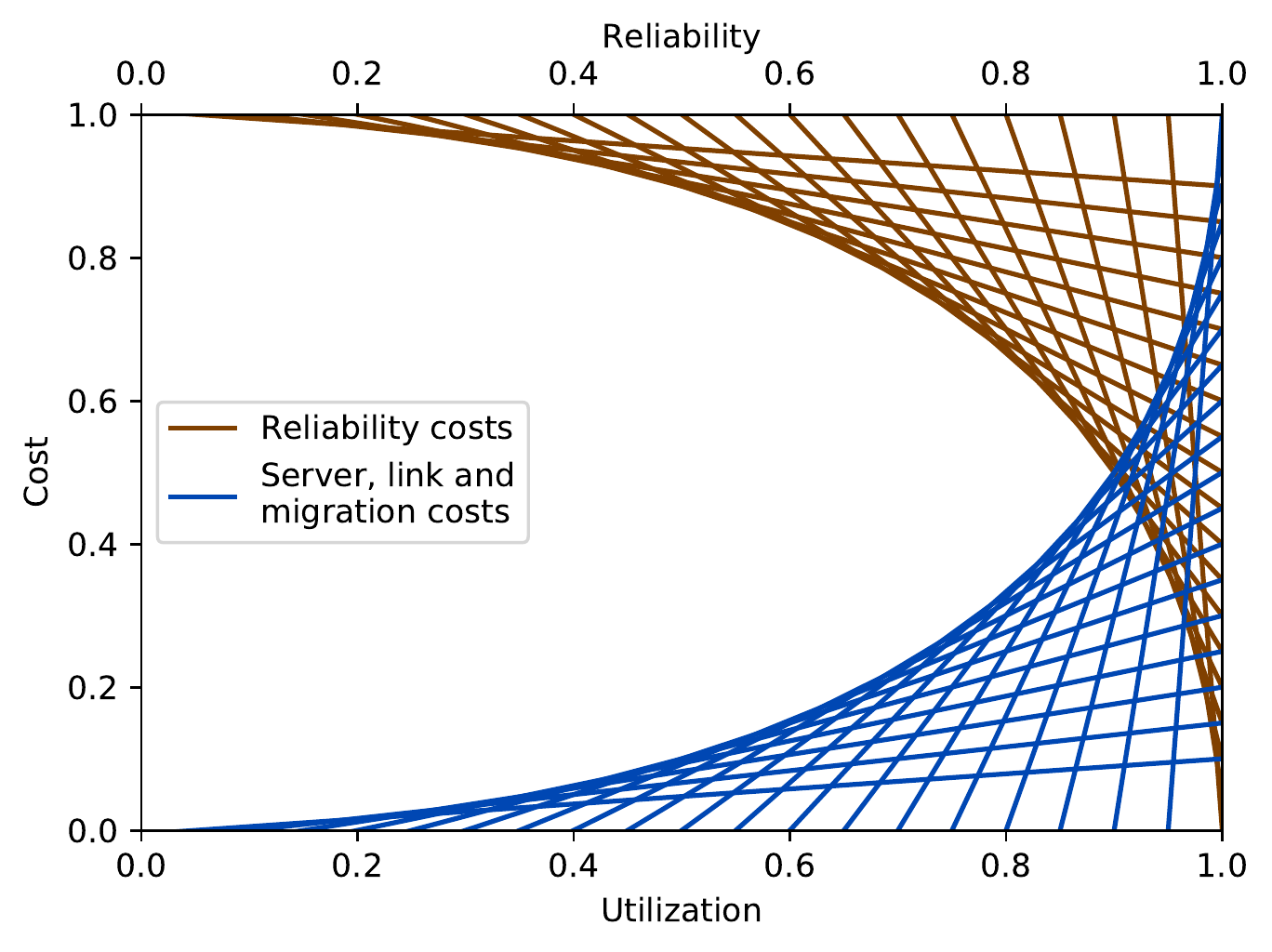}
	\caption{Penalty cost functions}
	\label{costs}
\vspace{-0.4cm}
\end{figure}

We now present the LP model with all the notation specified in Table \ref{parameters}. The objective function considers the \emph{minimization} of four different costs: reliability, server utilization, migration costs and link utilization i.e., 
\begin{multline} \label{objective_function}
	\frac{\alpha}{|S|} \sum_{s \in S} k_s + \frac{1 -\alpha}{|X|} \sum_{x \in X} k_x 
	+ \frac{1 - \alpha}{|V|} \sum_{s \in S} \sum_{v \in V_s} k_v + \frac{\beta}{|L|} \sum_{\ell \in L} k_\ell 
\end{multline}
the first three terms, i.e., reliability, server and migration costs, are weighted by the $\alpha$ parameter to enable the desired trade-off between the reliability of the services or the server load balancing and the number of migrations. While the link costs are in all cases taken into consideration, the small $\beta$ factor decreases its weight in comparison to others. It should be noted that all terms are normalized by the total number of services $|S|$, servers $|X|$, functions $|V|$ and links $|L|$, respectively. The reliability cost is defined by the resulting values from the piecewise linear cost functions $z_i(u) = 1 - (a_i u - b_i)$, corresponding to an inverse exponential cost, while the rest of costs follow $y_i(u) = a_i u - b_i$, that correspond to an exponential cost. Both linear cost functions are shown in Fig. \ref{costs}. It should be noted that the cost of a replica is implicitly included in the cost of the server where the replica is. This is because the cost to place a replica increases the server utilization due the overhead incurred by the creation of a new VNF. Therefore, the model will only do replication in case the benefit on either the server load balancing, or on reliability, compensates for the cost of creating new VNF replica.

Even though the reliability is considered as a parameter for this model, we are trying to optimize the placement of VNFs maximizing the achieved service reliability. For this reason, the previous equation (\ref{reliability_equation}) should be rewritten as:
\begin{equation} 
	\forall s \in S: R_s = \prod_{v \in V_s} \bigg[ 1 - \prod_{w \in V_{v}^s} \sum_{x \in X}(1 - R_x^w \cdot f_x^{w,s}) \bigg]
\end{equation}
, where the variable $f_x^{w,s}$ indicates if the function $w$ from service chain $s$ is allocated in server $x$. Note that, we assume the reliability of function $w$ is the same than the reliability of the physical server $x$ where the VNF is allocated. Because this solution is not linear, we propose to use reliability costs simulating an inverse penalty function based on the reliability of the server where the VNF is placed:
\begin{equation}
	\forall s \in S, \forall z \in Y: k_s \geq z \Big( \sum_{v \in V_s} \sum_{x \in X} f_x^{v,s} R_x\Big)
\end{equation}
, where $R_x$ is the reliability of the server $x$. Similarly, the server, migration and link costs are respectively defined by:
\begin{equation}
	\forall n \in N, \forall y \in Y: k_x \geq y \big( u_{x} \big)
\end{equation}
\begin{equation}
	\forall \ell \in L, \forall y \in Y: k_{\ell} \geq y \big( u_{\ell} \big)
\end{equation}
, but using the penalty costs that simulate the exponential function. The server utilization ($u_x$) is calculated by adding the utilization of every VNF, the overhead introduced by the creation of the VNF and the reserved capacity, in case of applying the N-to-N scheme. Therefore, $\forall x \in X$: 
\begin{equation}
     \sum_{s \in S} \sum_{v \in V_s}  \bigg[\sum_{\lambda \in \Lambda_s}  \frac{\lambda \cdot f_{x,\lambda}^{v,s} \cdot L_v}{C_x} \bigg] + \bigg[E_r u_v + \frac{f_{x}^{v,s}}{C_x  E_r} \bigg] + d_x^{v, s}\leq 1
\end{equation}
, where the variable $f_{x,\lambda}^{v,s}$ specifies if a certain traffic demand $\lambda$ is using the VNF $v$ in server $x$. If true, then using the corresponding load ratio $L_v$ for the specific function $v$, the bandwidth from $\lambda$ is added. To calculate the overhead, the first term adds the percentage that increases with the utilization of the VNF, and is pondered by the parameter $E_r$. The second term is also pondered by $E_r$ and adds a fixed minimum percentage that any VNF has due to its existence. The calculation of the reserved capacity in server $x$ for N-to-N is defined as:
\begin{equation}
\forall s \in S, \forall v \in V, \forall x, z \in X: d_x^{v, s} \geq \frac{1}{F_{MAX}} \sum_{\lambda \in \Lambda_s}  \frac{\lambda \cdot f_{z,\lambda}^{v,s} \cdot L_v}{C_z}
\end{equation}
, when $ x \neq z$. Therefore, the variable $d_x^{v, s}$ determines which is the replica with highest load and reserves a proportion in relation with the number of replicas. On the other hand, the link utilization is defined as:
\begin{equation}
    \forall \ell \in L: u_{\ell}   = \sum_{s \in  S} \sum_{\lambda \in  \Lambda_s} \sum_{p \in P_s} \frac{\lambda \cdot  t_{p}^{\lambda,s}  \cdot T_{p}^\ell}{C_\ell} \leq 1
\end{equation}
, where the variable $t_{p}^{\lambda,s}$ specifies when a specific traffic demand $\lambda$ is using the path $p$. If true, then the condition $T_{p}^\ell$ checks if path $p$ is traversing the link $\ell$, in order to sum the bandwidth $\lambda$ to the equation. 

Before optimizing the migration or replication, the model generally first optimizes the placement for each source-destination pair of nodes, with a specific ordered set of functions $V_s$ that belong to the service chain ${s \in S}$. In this case, the objective is only to load balance the server and link utilization and no replicas are allowed. In other words, the placement of VNFs determines the initial allocation of VNFs (i.e. $F_{x}^{v,s}$) that will be taken as input parameters for the replication and migration later. Taking this into account, the migration costs are defined by:
\begin{equation}
	\forall s \in S, \forall v \in V_s, \forall x \in X: k_v =  F_{x}^{v,s} (1 - f_{x}^{v,s}) E_m 
\end{equation}
, where the parameter $F_{x}^{v,s}$ specifies if the function $v$ was placed on server $x$ during the initial placement. If so, the variable $f_{x}^{v,s}$ determines if the function remains on the same server, which sets the cost to be zero, or the function has migrated, which sets the cost to be $E_m$.

When the model is not able to improve reliability by migrating functions, then it will try place replicas of VNFs. Then, the number of active paths for each service chain is related to the possible allocation of VNFs in the servers and, therefore, is constrained by the number of replicas (i.e. $F_{MAX} \geq$ 1). Therefore, 
\begin{equation} \label{rmax}
	\forall s \in S: 1 \leq \sum_{p\in P_s} t_{p}^s \leq F_{MAX} + 1
\end{equation}
How many times a VNF can be replicated is determined by the parameter $F_v$ and constrained by: 
\begin{equation}
\forall s \in S, \forall v \in V_s:  \sum_{x \in X} f_x^{v,s} \leq F_v \sum_{p \in P_s} t_{p}^s + 1 - F_v
\end{equation}
If the function can be replicated, then the function can be placed in many servers as active paths the variable $t_{p}^s$ determines. The following routing constraint defines that each traffic demand $\lambda$ from each service chain $s$ can only use one path $p$:
\begin{equation}
\forall s \in  S, \forall \lambda \in  \Lambda_s: \sum_{p \in P_s} t_{p}^{\lambda,s} = 1
\end{equation}
Then, when some traffic demand $\lambda$ is using the path $p$, this is activated for the service chain $s$:
\begin{equation} \label{routing_1}
\forall s \in  S, \forall \lambda \in  \Lambda_s, \forall p \in P_s : t_{p}^{\lambda, s} \leq t_{p}^{s} \leq \sum_{\lambda' \in \Lambda_s} t_{p}^{\lambda', s}
\end{equation}
Then, the next constraint allocates all VNFs from the service chain $s$ in the activated path using the variable $f_{x,\lambda}^{v,s}$:
\begin{equation}
\forall s \in  S, \forall p \in P_s, \forall \lambda \in \Lambda_s, \forall v \in  V_s: t_{p}^{\lambda, s} \leq \sum_{n \in N_p} f_{x,\lambda}^{v,s} 
\end{equation}
, where $N_p$ is an ordered set of servers traversed by path $p$.

The rest of constraints assure the proper activation of VNFs in the correct order for each traffic demand. First, the next constraint specifies that each traffic demand $\lambda$ has to traverse an specific function $v$ in only one server:
\begin{equation}
\forall s \in S, \forall v \in  V_s, \forall \lambda \in \Lambda_s: \sum_{x \in  X} f_{x,\lambda}^{v,s} = 1
\end{equation}
Then, the next constraint allocates the function $v$ on server $x$ when at least one traffic demand is using it:
\begin{equation}
\forall s \in  S, \forall v \in  V_s, \forall x \in X, \forall \lambda \in \Lambda_s: f_{x,\lambda}^{v,s} \leq f_x^{v,s} \leq \sum_{\lambda' \in  \Lambda_s} f_{x,\lambda'}^{v,s} 
\end{equation}

Finally, since each service chain is composed by a certain ordered set of VNFs, each traffic demand has to traverse them in the correct order, i.e.,
\begin{multline}
	\forall s \in S, \forall \lambda \in \Lambda_s, \forall p \in P_s, \forall v \in {V_s}, \forall n \in N_p, \forall x \in X_n: \\
	\Bigg( \sum_{m = 0}^{n} \sum_{y \in X_m} f_{y, \lambda}^{(v-1),s} \Bigg) - f_{x, \lambda}^{v,s} \geq t_{p}^{\lambda,s}  - 1 \quad if \quad v>0
\end{multline}

, where for every traffic demand $\lambda$, the function $v$ is allocated at server $x$, only if the previous function $v-1$ of the same service chain $s$ is already allocated in any of the previous available servers $y$ from the activated path $p$. Note that $N_p$ is an ordered set of nodes traversed by the path $p$ and $X_n$ is the set of servers running on node $n$. 

\begin{figure}[!t]
	\centering
	\includegraphics[width=0.9\columnwidth]{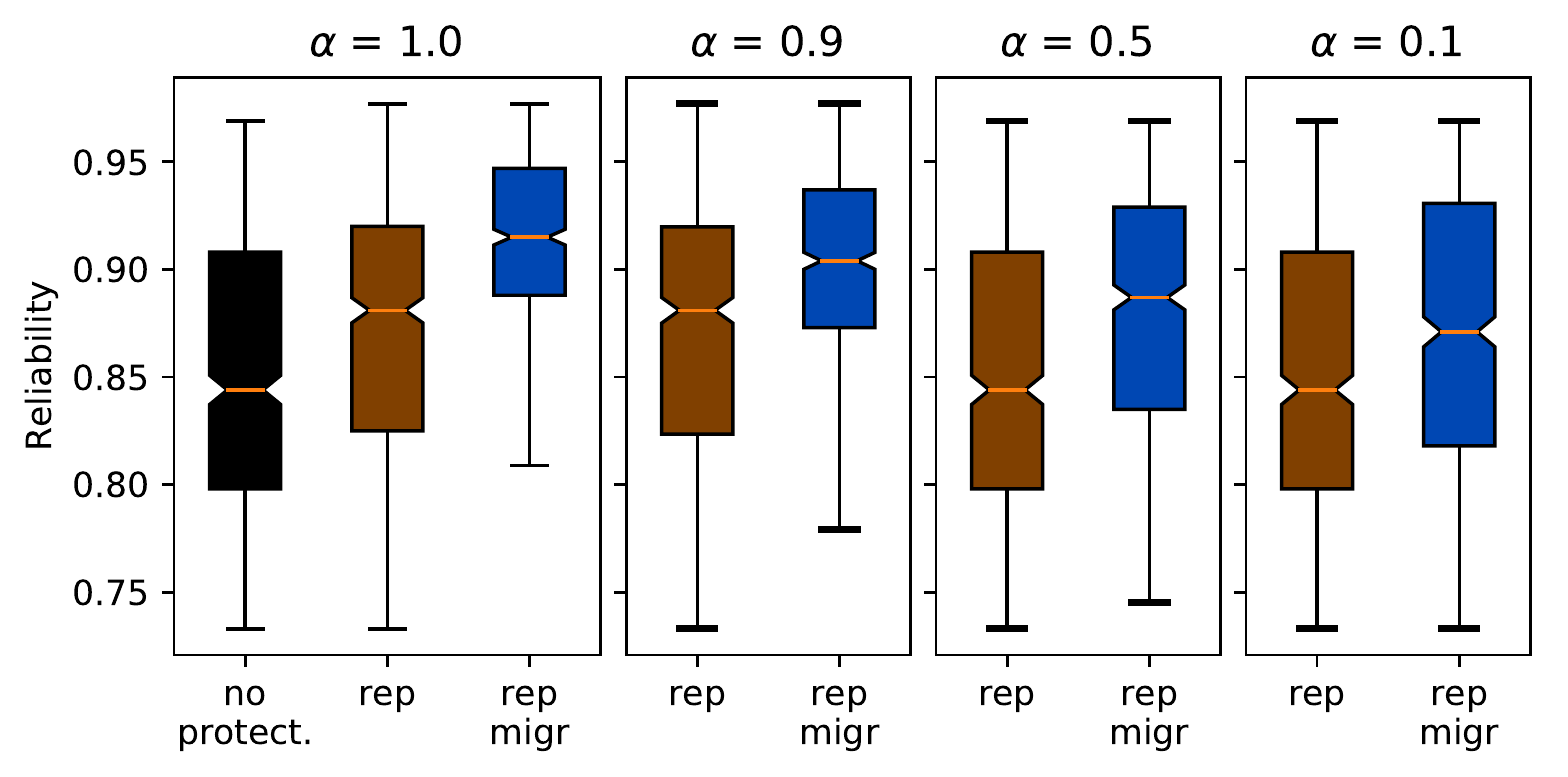}
	\caption{Reliability}
	\label{reliability}
\vspace{-0.4cm}
\end{figure}
\begin{figure*}[!t]
	\centering
	\includegraphics[width=0.8\textwidth]{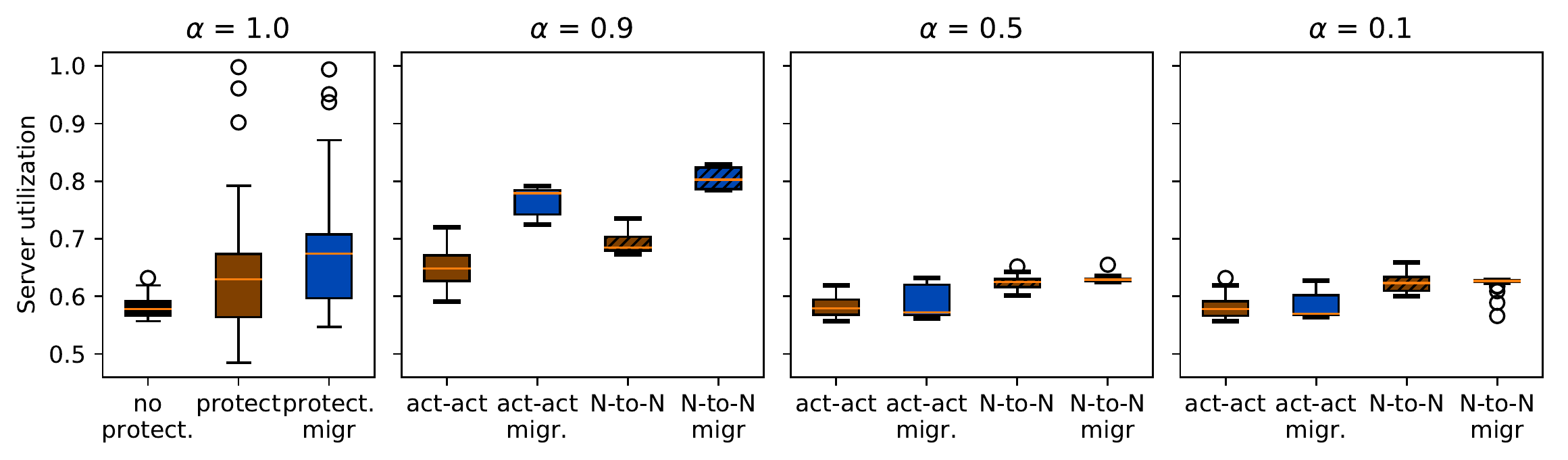}
	\caption{Server utilization}
	\label{server}
\vspace{-0.4cm}
\end{figure*}

\section{Performance Evaluation}

This section shows the results from the LP model implemented with Gurobi Optimizer \cite{gurobi}. The analyzed topology is \emph{janos-us} with 26 nodes and 84 links taken from SNDLib \cite{sndlib}. Each node has one PM with 2000 units of capacity to allocate VNFs, and the capacity of each link is 1 Gbps. Each source-destination pair of nodes allocate one service chain and randomly generates between 1 and 6 connections with a random bandwidth between 1 and 10 Mbps. Each service chain is composed by 3 VNFs (1950 VNFs in total) and the maximum number of replicas per service chain is 5. Each PM has a random reliability between 0.9 and 0.99.

\begin{table}
\centering
\caption{Number of migration and replications}
\label{replicas_migrations}
    \begin{tabular}{c|c|c|c|c}
    ~         & $\alpha$ = 1.0 & $\alpha$ = 0.9 &  $\alpha$ = 0.5 & $\alpha$ = 0.1 \\ \hline
    rep       & 470   & 472   & 15    & 8  \\ \hline
    rep\_migr & 1165-1198  & 1098-1193   & 1425-54  & 1420-1   \\ \hline
    \end{tabular}
\end{table}

\subsection{Reliability}

Fig. \ref{reliability} shows the results on reliability when no protection scheme is applied (i.e. \emph{no\_protect.}), when only replicas (active-active or N-to-N, indistinctly) are allowed for protection (i.e. \emph{rep}) and when, both, replications and migrations are allowed (i.e. \emph{rep\_migr}). Because both, active-active and N-to-N, protection schemes have the same benefits on the achieved reliability getting similar results, both are represented under the same plot due to space limitations. Different results are shown depending on the chosen $\alpha$ value specified in the objective function (\ref{objective_function}). 

Without protection, $\alpha$ does not affect the results because neither replicas nor migrations are allowed. When the model only optimizes reliability (i.e. $\alpha$ = 1), only placing 470 replicas (see Table \ref{replicas_migrations}) increases the reliability, but even more evident is this gain when 1198 migrations and 1165 replications are combined. For the rest of $\alpha$ values, doing replications only have benefit when $\alpha$ = 0.9. Contrary, when replications and migrations are combined, the benefit on the achieved reliability is more relevant. Note that, when the model is able to optimally reallocate VNFs, there are more released resources from critical nodes and, therefore, more chances to allocate replicas. The results also show how the increment on the number of replicas to provide protection improves the reliability, but when this is combined whit an optimum reallocation of VNFs the benefit is clearly higher.

\begin{figure}[!t]
	\centering
	\includegraphics[width=0.9\columnwidth]{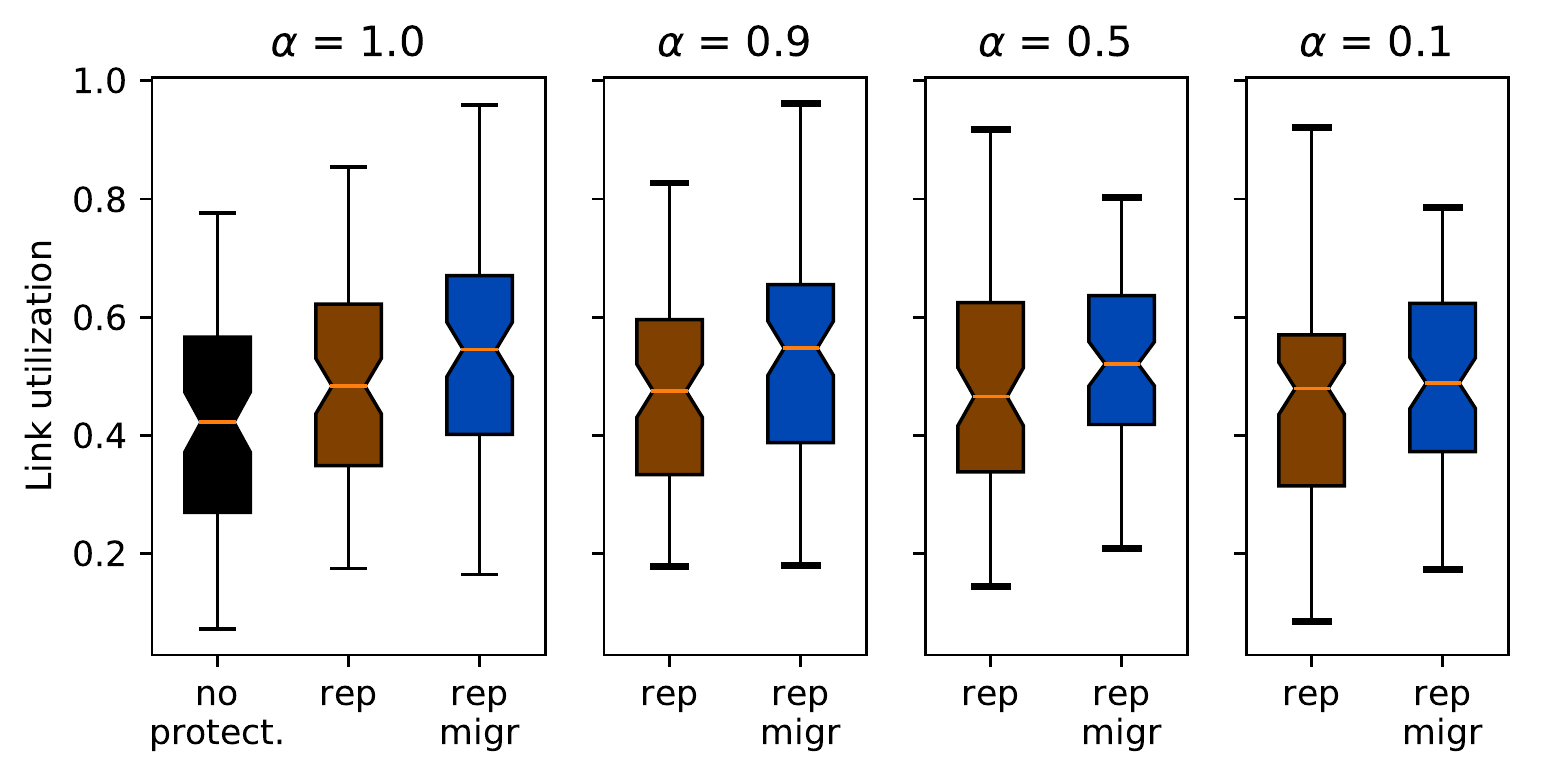}
	\caption{Link Utilization}
	\label{link}
\vspace{-0.4cm}
\end{figure}

\subsection{Server and link utilization}

To analyze the servers load balancing, the Fig. \ref{server} show the utilization of all servers in the network for different $\alpha$ values and protection schemes. When $\alpha$ = 1, allowing replication jointly with migrations (i.e. \emph{protect.\_migr.}) implies a higher server utilization due to the higher number of replicas compared to the case where no migrations are allowed (i.e. \emph{protect.}). To be noted, that in both cases there are servers completely overloaded. By relaxing $\alpha$ to 0.9, clearly avoid to have overloaded servers, and all schemes are able to load balance all servers utilization. In this case, having migrations increases the average utilization compared to the case where only replication is allowed. For the rest of $\alpha$ values the behavior is comparable but decreasing the average utilization even more. To be noted that N-to-N schemes increase the average server utilization, but taking into account their benefits on fast recovering, this increment could be justified.

Fig. \ref{link} shows the link utilization, one more time, for different $\alpha$ values and compares the case when only replicas are allowed (i.e. \emph{rep}) and when replicas and migration are allowed (i.e. \emph{rep\_migr.}). Here, only performing replications increases the average link utilization in all cases. If re-allocations are allowed by doing migrations, then, the average is also increased in all cases, specially critical when $\alpha$ is close to 1 creating bottlenecks in some links. These results show how a good balance between reliability and load balancing (i.e. $\alpha$ = 0.5) is important to maintain the network load balanced.

\section{Conclusions}

To provide high reliability in service function chains, a high number of VNF replicas is required. While this increment becomes necessary to meet the reliability requirements, the utilization of server and network resources is affected. We proposed to use migrations of VNFs jointly with active-active configurations to reduce the number of replicas while maintaining an acceptable level of reliability. We also showed the impact on the server utilization of N-to-N configurations, that is comparable higher than active-active but benefiting from fast recovery.

\section*{Acknowledgment}
This work has been performed in the framework of SENDATE-PLANETS (Project ID C2015/3-1), and it is partly funded by the German BMBF (Project ID 16KIS0470).

\bibliographystyle{IEEEtran}
\bibliography{ICC18}

\end{document}